\documentclass[letterpaper]{article}
\usepackage{natbib,alifeconf}  
\usepackage{amssymb}

\usepackage[hidelinks]{hyperref}

\usepackage{booktabs}

\newcommand\blfootnote[1]{%
  \begingroup
  \renewcommand\thefootnote{}\footnote{#1}%
  \addtocounter{footnote}{-1}%
  \endgroup
}

\makeatletter
\renewcommand{\paragraph}{%
  \@startsection{paragraph}{4}%
  {\z@}{1.75ex \@plus 1ex \@minus .2ex}{-1em}%
  {\normalfont\normalsize\bfseries}%
}
\makeatother

\usepackage{bm}

\usepackage{comment}
\usepackage{color}
\usepackage{nicefrac}

\usepackage{optidef}

\usepackage{stackengine}

\usepackage[caption=false,font=normalsize]{subfig}

\usepackage{todonotes}
\usepackage[shortlabels]{enumitem}

\definecolor{cambridgeblue}{rgb}{0.64, 0.76, 0.68}

\usepackage{etoolbox}
\usepackage{url,cleveref}
\newtoggle{ShowNames}
\toggletrue{ShowNames}

\title{Parasitic Masquerade: Societal Scale Human--Machine Interaction}

\author{
    Jiejun Hu-Bolz$^{1}$ \and
    James Stovold$^2$ \\
    \mbox{}\\
    $^1$Lancaster University Leipzig, Germany \\
    $^2$University of York, UK \\
    \url{j.hu14@lancaster.ac.uk}
}

\begin{document}
\maketitle

\begin{abstract}
This work extends recent developments in studying human--machine interaction by scaling from individual game-theoretic models to a societal-level model. We adopt a Graphon Mean-Field Game (GMFG) that models the interaction among four groups of internally-homogeneous but externally-heterogeneous agents in a shared environment. Our results show that parasitism can masquerade as productive learning, with knowledge distribution and actions appearing healthy while being driven by machine coupling rather than independent investigation. To detect this, we measure the direction of information flow and belief entropy of the environment, revealing that human to machine channel dominates across all scenarios, with the asymmetry intensifying under parasitism. We further demonstrate that the system exhibits coexisting mutualistic and parasitic equilibria, where environmental noise can induce a tipping point that shifts agents past the cognitive cost barrier. These emergent phenomena are not designed into any individual agent but arise from the collective interaction structure, underscoring the need to study the sociology of humans and machines holistically as a complex system.


\end{abstract}

\noindent{}Data/Code available at: \url{https://github.com/JunHu-Bolz/HMSS_GMFG_ALIFE26.git}
\blfootnote{\textcopyright 2026 J.\ Hu-Bolz \& J.\ Stovold. Published under a Creative Commons Attribution 4.0 
International (CC BY 4.0) license.}

\section{Introduction}
Human--machine interaction has reached a point where its effects are no longer confined to cyberspace but permeate the physical world. Take OpenClaw as an example. As one of the fastest-growing open-source AI projects in GitHub's history, it gained rapid popularity as an autonomous AI assistant capable of coding, task automation, and tool integration. Its user base is notably heterogeneous: technically proficient users such as developers tend to scrutinise system behaviour closely, while others rely on its outputs uncritically, sometimes disregarding contextual information altogether. Given the scale of adoption, the heterogeneity in how users engage with the system creates a distribution of trust states across the population, which is hard to manage and potentially dangerous, such as misinformation and polarisation.

Machines equipped with various fundamental models and behavioural preferences also interact at scale. Machine-to-machine interaction operates at speeds far beyond human oversight, meaning misaligned behaviours can compound before any intervention is possible (e.g., Flash Crash). Critically, individual machines may each behave correctly with respect to their own objectives while their interaction produces collectively misaligned outcomes. Recursive feedback loops further amplify this risk, as machines can act on each other's outputs repeatedly and rapidly in ways that human populations, constrained by cognitive bandwidth and reaction time, typically cannot. This is a classic example of a `wicked problem'~\citep{conklin_wickedproblemssocial}, and so the best way to understand large-scale human--machine interaction is with a holistic approach~\citep{tsvetkova2024new, hu2024human}.

We strive to provide a quantitative analysis of human-machine interaction. In \citep{hu2024human}, we adopt mean-field game to study how human satisfaction shapes the cooperation dynamics and intelligence maturity of large populations of interconnected autonomous machines in a Human-Machine Social System \citep[HMSS;][]{tsvetkova2024new}, finding that high satisfaction promotes machine knowledge growth but can also trigger disruptive, such as non-cooperative machine behaviour. We then focus on human's decision-making process in~\citep{hu2025can}, where we modelled the interaction among environment, human, and machine as a stochastic game, capturing how parasitic relationship emerges between a single human and a single machine through the lens of information theory. We demonstrated that factors such as initial trust, early positive experience, and environmental dynamics collectively drive over-reliance. 

\begin{figure}[h]
    \centering
    \includegraphics[width=1\linewidth]{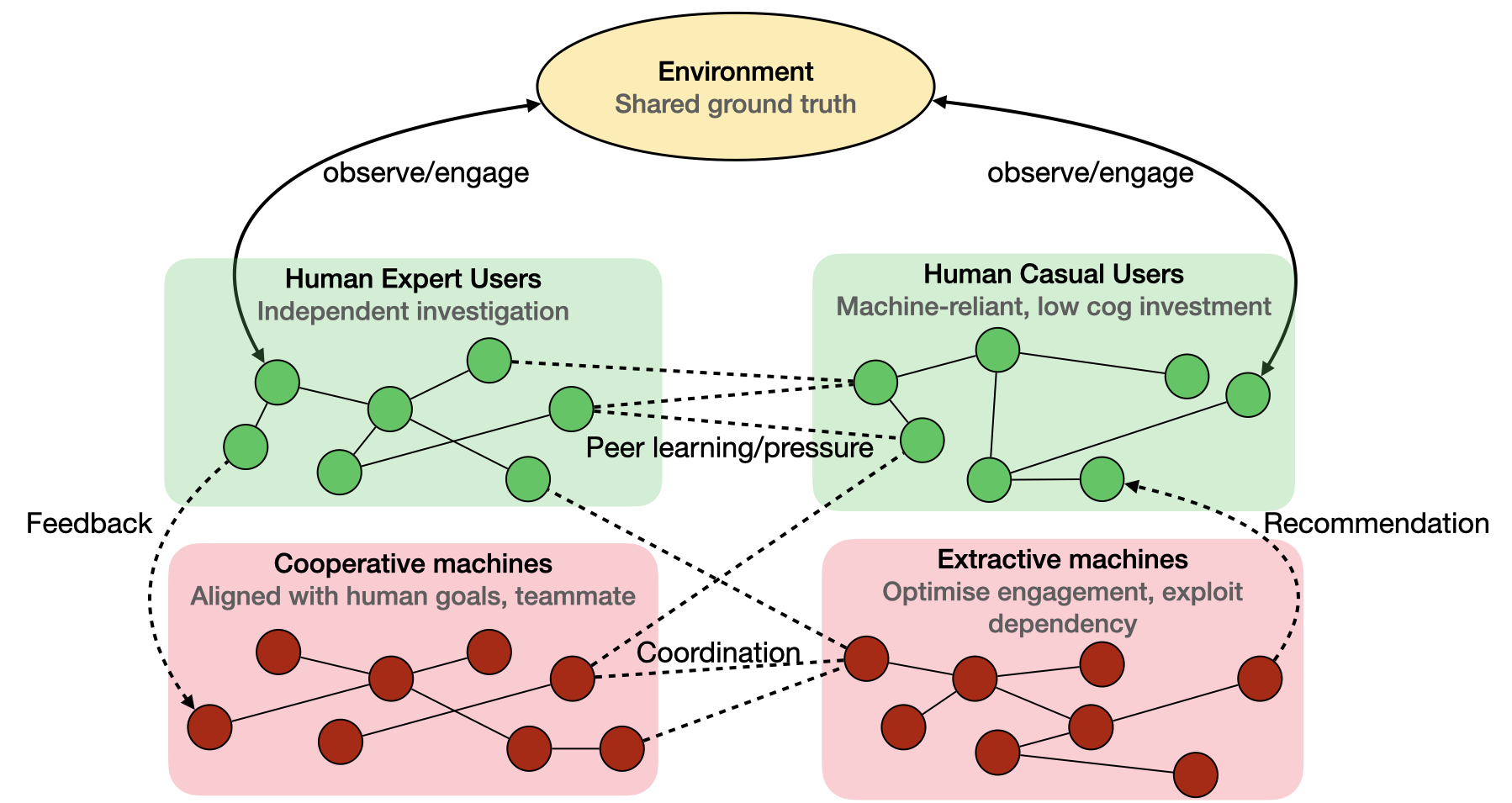}
    \caption{Human-Machine Social System: In HMSS, there exists different groups of human and machine agents. They interact freely. Only human agents can observe and engage with this environment directly; machines obtain in-situ information only through interaction with humans. The population is divided into four groups: human expert users, human casual users, cooperative machines, and extractive machines. Human-human interaction includes peer learning or pressure. Human-Machine interaction includes human obtaining recommendation from machines or providing feedback to the machines. Machine-machine interaction includes coordination and learning. }
    \label{fig: intro}
\end{figure}

In this work, we aim to gauge the gap between the microscopic view (one human--machine pair), and the macroscopic view (societal phenomena of populations of heterogenous human and machine agents, as shown in Fig.~\ref{fig: intro}). 
A classic mean-field game (MFG)~\citep{lasry2007mean} models strategic interaction of infinitely many rational agents, where each agent optimises its own cost while being influenced by the aggregated behaviour of the population through a single mean-field term. This reduces an intractable $N$-player game to a representative agent problem coupled with a population distribution. MFG, however, is not sufficient to fully capture all human--human, human--machine, and machine--machine interactions in HMSS (Fig.~\ref{fig: intro}), since it assumes all agents are interchangeable, i.e., each agent sees the same mean-field.  

A Graphon Mean-Field Game (GMFG)~\citep{caines2021graphon} relaxes this assumption by introducing a graphon kernel. 
A graphon~\citep{lovasz2006limits} is a symmetric measurable function that arises as the limit of sequences of dense graphs, providing a continuous representation of interaction structure in large populations. In GMFG, it specifies the coupling weight between any agents, so that each agent receives distinct weighted mixture of the population's mean-field signals. 
This allows the framework to represent a distribution of agent types and their interaction patterns simultaneously, capturing population heterogeneity without sacrificing the computational tractability that makes MFG attractive in the first place. We investigate three research questions:
\begin{itemize}[noitemsep]
    \item What unexpected behaviour arises under different interaction regimes in HMSS? We study two regimes, i.e., mutualistic, parasitic, each in a baseline form and with collusion machines (four scenarios). We demonstrate the terminal knowledge distributions, information flow, and parasitism index that distinguish the two regimes, and show that parasitism can masquerade as productive learning. 
    \item How does information flow between humans and machines, and when does it become extractive?  We measure causal information flow in each direction and show that the human-to-machine channel dominates across all scenarios, with the asymmetry intensifying under parasitism.
    \item Does the system admit multiple equilibria, and what determines which one is selected? We show that the GMFG admits coexisting mutualistic and parasitic equilibria under identical parameters, with the outcome determined by the interaction between information noise and the initial knowledge distribution.  A noise-induced tipping point at a critical threshold destroys the mutualistic equilibrium via a positive feedback cascade through the cognition cost barrier. 
\end{itemize}

\section{Related Work}
Recent work argues that the next phase of AI will manifest not as monolithic superintelligence but as distributed, multi-agent ecosystems where human and AI co-evolve through social and relational dynamics~\citep{evans2026agentic}. It is challenging to conceptually capture the interaction of heterogeneous agents at scale. Additionally, with the rapid development of AI, high-quality datasets are delayed and sparse, which hinders empirical study. This makes theoretical models (such as this work) valuable as tools to understand such human--machine complex system, answering the call of~\citet{pedreschi2025human}. 

Despite attempts to understand human--AI interaction at scale by formalising it as an optimisation problem~\citep{nayebi2026intrinsic}, we firmly believe that the environment where the agents are located is essential to the outcomes of the interaction as demonstrated in our previous work~\citep{hu2025can}. This previous work focussed on the scenario of one human and one machine (similar to most Human--AI interaction research, e.g.~\citep{alvarez_fosteringcreativitymixed, walton_evaluatingmixedinitiative, walton_frommetricsto}). By ensuring the environment is taken into account, we follow principles of embodied and situated cognition~\citep{varela_embodiedmindcognitive} where the cognitive agent interacts with itself and the environment through its body and perceptive world view. While some work has been completed on measuring the level of distraction while playing games~\citep{cutting_investigatinggameattention}, there is limited substantive research that accounts for the wider context in which a task takes place (although there have been calls for this to change~\citep{vanhove_humancomputerinteraction}). To capture this heterogeneity and the role of the environment together, we adopt a Graphon Mean-Field Game to analyse human--AI interaction at scale.

\begin{table}[h!]
\centering
\caption{Notation and parameter ranges}\label{tab: notation}
\setlength{\tabcolsep}{3pt}
{\small
\begin{tabular}{@{}clc@{}}
\toprule
Symbol & Description & Range / Value \\
\midrule
\multicolumn{3}{@{}l}{\textit{Indices and spaces}} \\[2pt]
$i$         & Generic agent (either type)        & --- \\
$h$, $m$    & Human / machine agent              & --- \\
$\mathcal{H,M}$  & Set of human / machine agents & --- \\
$j$, $k$    & Group index                        & --- \\
$s^h$         & Human knowledge state               & $[0, 1]$ \\
$a^h$         & Human action               & --- \\
$c^h(s^h)$      & Human cognition cost     & --- \\
$s^m$         & Machine cooperative state               & $[0, 1]$ \\
$a^m$         & Machine action               & --- \\
$e$         & Environment state                  & $[0, 1]$ \\
$\mu^{h,m}_k(t)$ & Mean-field term of group $k$  & --- \\
$\rho_k(t,s)$ & Probability density of group $k$ & --- \\
$b(e)$      & Belief of the environment & --- \\
$\mathcal{N}(i)$    & Agent $i$'s group  & --- \\
$\mathcal{N}_\Omega$  &  Other groups without agent $i$ & --- \\
\midrule
\multicolumn{3}{@{}l}{\textit{Drift parameters}} \\[2pt]
$A_{h,m}$   & Action sensitivity                 & $0.3$ \\
$B_{h,m}$   & Mean-reversion strength            & $[-0.5, -0.3]$ \\
$C_{h,m}$   & Intra-group coupling weight      & $[0.05, 0.30]$ \\
$D_{h,m}$   & Inter-group coupling weight      & $[0.20, 1.20]$ \\
$\sigma_{h,m}$          & Diffusion (noise)        & $[0.10, 0.233]$ \\
\midrule
\multicolumn{3}{@{}l}{\textit{Cost parameters}} \\[2pt]
$\gamma_{h,m}$          & Group deviation penalty  & $[0.9, 4.5]$ \\
$\gamma_{h,m}^{\times}$ & Cross-group alignment  & $[0.9, 6.0]$ \\
$\alpha_{h}$          & State reward             & $[0.9, 4.5]$ \\
$\kappa$              & Cognition cost scale     & $\{0.7, 1.0\}$ \\
\midrule
\multicolumn{3}{@{}l}{\textit{Coupling}} \\[2pt]
$w_{hh}$ & $h$--$h$ graphon weight          & $[0.20, 0.50]$ \\
$w_{hm}$ & $h \leftarrow m$ graphon weight  & $[0.10, 0.70]$ \\
$w_{mh}$ & $m \leftarrow h$ graphon weight  & $[0.20, 0.50]$ \\
$w_{mm}$ & $m$--$m$ graphon weight           & $[0.05, 1.00]$ \\
$Z_{hh}$, $Z_{hm}$  & Intra-/inter- group coupling for $h$ & --- \\
$Z_{mm}$, $Z_{mh}$  & Intra-/inter-group coupling for $m$ & --- \\
\midrule
\multicolumn{3}{@{}l}{\textit{Solver}} \\[2pt]
$N_s$       & State grid points                  & $101$ \\
$N_t$       & Time steps                         & $200$ \\
$N_e$       & Environment grid points            & $50$ \\
\bottomrule
\end{tabular}
}
\end{table}
\section{Modelling HMSS with GMFG}
In the proposed model, we illustrate large-scale human--machine interaction through the example of a human interacting with AI. Due to the complexity of the proposed model, we refer the readers to Table~\ref{tab: notation} for key notations, their meanings, and the parameter ranges.  

Let each agent be characterized by a type that determines their position in the interaction network. The human population occupies types $h$ and the machine population types $m$. 
Compared to a traditional mean field game where all neighbours are interchangeable (their influence is captured by one mean-field term), we define the agent's specific view when interacting with humans and machines by using a graphon. 
There are four types of interaction among human and machine agents via graphon kernel $w$: human-human social interaction where other humans' states affect one agent's state, $w_{hh}$; Human-machine interaction where human's prompt influence the machine's state, $w_{hm}$; Machine-human interaction where machine's action affects human's state, $w_{mh}$; machine-machine interaction where machines share information, $w_{mm}$. 

We assume a confined environment with a common task for all the human agents, the machines do not have agency to solve the task directly, the only way for the machines to obtain in-situ information is through interaction with the human agents. The $h$-agents perceive the environmental state, $e \in \mathcal{E}$, forming the state $s^h(e)$, where $h \in \mathcal{H}$, a function of the environmental state. We can define the cognitive cost as a function of perceiving the environment, $c^h (s^{h})$. Human agents act, $a^h$, to engage with the environment based on the knowledge state, $s^h$. The action is eventually composed into prompts to solve the task. For machines, we define the state, $s^m$ where $m \in \mathcal{M}$, as the cooperation level. We then define the machine's action, $a^m$, as the effort in producing high-quality output. We denote $\rho_k(t,s)$ as the probability density of group $k$ over the knowledge state space, from which the mean-field term is computed as $\mu_k(t) = \int_0^1 \rho_k(t,s)\ ds$.

We distinguish two levels of environmental knowledge. An observation of the environment is a momentary, noisy signal that a human agent obtains by investigating the environment, whose quality depends on the agent's knowledge state $s^h$ based on the environment state and action $a^h$. A belief $b_k(e)$ is the accumulated probability distribution over the true environmental state for group $k$, updated over time as observations are integrated:
\begin{equation}
    b_k^{n+1}(e) = (1 - r_s - r_d) b_k^n(e) + r_s b^*(e) + r_d b_{\text{uni}}(e)
\end{equation}
where $r_s = \lambda_k \int a^h(t, s) \rho_k(t,s)\ ds$ is the rate of belief sharpening toward the truth, proportional to the population-weighted mean action and a group-specific learning rate $\lambda_k$, $r_d$ is the rate of belief drifting to ignorance, $b^*(e)$ is the target belief (perfect knowledge), and $b_{\text{uni}}(e) = 1/N_e$ is the uniform belief (no knowledge). Higher action magnitude sharpens the belief toward the true state, while inactivity allows the belief to diffuse back toward the uniform prior. We measure the uncertainty of group $k$'s belief by the Shannon entropy $H_k(E) = -\sum_{e \in \mathcal{E}} b_k(e) \ln b_k(e)$, which decreases as the belief concentrates near the ground truth. 

\paragraph{GMFG Preliminaries}
In GMFG, it considers evenly distributed finite groups of agents. The group $\mathcal{N}$ for agent~$i$ (h-type or m-type in HMSS) is denoted $\mathcal{N}(i)$. Please note we denote agent $i$ in the general case of GMFG. The averaged state dynamics and cost can be decomposed into contributions from the sub-groups connected to $i$ through the graphon kernel. 

The state dynamics of agent $i$ is influenced within the group by its neighbours (local influence) and by other groups connected to it, $\mathcal{N}_\Omega$, (global influence). We can define the local, $f_0$ and global, $f_\Omega$, dynamics as 

{\small
\begin{align}
    &f_0 (s_i, a_i, \mathcal{N}(i)) = \frac{1}{|\mathcal{N}(i)|}\sum_{q \in \mathcal{N}(i)} f_0 (s_i, a_i, s_q) \label{eq: local}\\
    &f_\Omega (s_i, a_i, g_{\mathcal{N}(i)}) =\frac{1}{M} \sum_{\Omega=1}^M g_{\mathcal{N}(i)\mathcal{N}_\Omega} \frac{1}{|\mathcal{N}_\Omega|}\sum_{q \in \mathcal{N}_\Omega} f(s_i, a_i, s_q) \label{eq: global}
\end{align}}

\noindent{}$f_0$ is the pairwise interaction function, which describes how agent $q$'s (in the same group as $i$) state affect agent $i$'s state. $f$ is the pairwise interaction for cross-group coupling. $M$ is the number of groups in graph and $g_{\mathcal{N}(i)\mathcal{N}_\Omega}$ is the graphon between $\mathcal{N}(i)$ and $\mathcal{N}_\Omega$. The dynamics of agent $i$ is $\tilde{f}(s_i, a_i, g_{\mathcal{N}(i)}) = f_0 + f_\Omega$. We then have the dynamics of agent's state $s_i(t)$ defined as a Stochastic Differential Equation (SDE)
\begin{equation}
    d s_i(t) = \tilde{f}(s_i, a_i, g_{\mathcal{N}(i)}) dt + \sigma_i d \mathcal{W}
    \label{eq: dyn_state}
\end{equation}
We capture the randomness by using Brownian motion, where $\sigma_i$ is the diffusion constant and $\mathcal{W}$ is a standard Wiener process. 
\paragraph{GMFG for HMSS}
In the proposed GMFG for HMSS, we use superscripts $h$ and $m$ to distinguish human and machine agents respectively, e.g.\ $s^h$ for the state of an $h$-agent, resp.\ for $m$. We assume that agents within a group are homogeneous. As the number of groups and the number of agents in each group tend to infinity, these summed neighbour contributions are represented by graphon integrals over the local mean fields of adjacent agents. 
\begin{align}
&f_0[s^h,a^h,\mu_t^h]
= \int_{\mathbb{R}} \left(f_0\!\left(s^h,a^h,z\right)\,\mu^h \right) dz, \\
&f_g[s^h,a^h, \mu^G ; w_h]
= \nonumber\\
&\quad \quad \quad \quad \int_0^1\int_{\mathbb{R}} \left(f_g\!\left(s^h,a^h,z\right)\,w_h\mu^G\right) dz\ dG
\end{align}
where $z$ is the state of another agent in the same group (as $s_q$ in Eq.~(\ref{eq: local})); $w_h$ is the graphon related to $h$-agent; $\mu^G$ is the ensemble of local mean fields, $\mu^h$ and $\mu^m$, where $G = \{h, m\}$~\citep{caines2021graphon}.

$h$-agents aim to reduce the information entropy in to a certain level in order to solve the problem. Therefore, we define the drift function for  $h$-agent in group $j$ considering their action, $h$-agents within the same group, $h$-agent's belief of the environment in group $k (k\neq j)$, and $m$-agents response in Eq.~(\ref{eq: drift_h}). For $m$-agents, the definition of the drift function is symmetrical as the $h$-agents.

{\small
\setlength\abovedisplayskip{-10pt}
\setlength\belowdisplayskip{6pt}
\begin{align}
    \tilde{f}_h &= A_h a^{h} + B_h (s^{h}-\mu^h) + C_h Z_{hh(j, k)} + D_h Z_{hm(j, k)} 
    \label{eq: drift_h} \\
    \tilde{f}_m &= A_m a^{m} + B_m (s^{m}-\mu^m) + C_m Z_{mm(j, k)} + D_m Z_{m h(j, k)} 
    \label{eq: drift_m}
\end{align}
}

\noindent{}$Z_{hh(j, k)}$ is the influence on an $h$-agent within group $j$ and other groups $k$, which captures the impact of other humans' belief on the current human's knowledge. $Z_{hm(j, k)}$ is the influence on an $h$-agent in group $j$ from $m$-agent in group $k$. $Z_{mj(j, k)}$ is the influence on an $m$-agent in group $j$ from $h$-agents in group $k$. Finally, $Z_{mm(j, k)}$ in the influence on an $m$-agent from other $m$-agents. They are defined as follows.
\begin{align}
\small
    Z_{hh(j,k)} &= \int_H w_{hh}(j,k) \int_E s^h(e) b_k(e,t) de\ dk \nonumber\\
Z_{hm(j,k)} &= \int_M w_{hm}(j,k)\mu^m_k(t) dk \nonumber\\
Z_{mh(j,k)} &= \int_H w_{mh}(j,k)\mu^h_k(t) dk \nonumber\\
Z_{mm(j,k)} &= \int_M w_{mm}(j,k) \mu^m_k(t)dk \nonumber
\end{align}
where $\mu^h$ and $\mu^m$ are local mean field term of $h$ and $m$ groups; $A_{h, m}$, $B_{h, m}$, $C_{h, m}$, and $D_{h, m}$ are weights.

According to the $s^h$, the action of $h$-agent leads to cognition cost $c^h$. They interact with their $h$-agent neighbours seeking for confirmation of understanding the environment (homophily) and $m$-agent for recommendation. Additionally, $h$-agents will be rewarded for reducing the environmental uncertainty by reaching a higher knowledge state (which can also be interpreted as satisfaction). Hence, we define the cost function of the $h$-agent as 
\begin{equation}
\small
L_h = \frac{1}{2}{a^h}^2 c^h + \gamma_h [s^h - (\mu^h+Z_{h h (k)})]^2 + \gamma_{h m} (s^h-Z_{h m (k)})^2 - \alpha^hs^h
\label{eq: cost_alpha}
\end{equation}
where the quadratic form of action can ensure diminishing returns. The cognitive cost is defined as $c^h(s^h)=e^{-\kappa s^h}$, where $\kappa$ is the cognition cost scale. This means that $h$-agents with high knowledge states incur a lower cognitive cost, and vice versa. The cognitive cost creates a barrier where low-knowledge $h$-agents face high cognitive costs, suppressing their action and preventing knowledge growth.
For $m$-agents (machines), the state $s^m$ is based on the foundation model, which leads to action $a^m(s^m)$. $m$-agents also interact with machines within the group and across $m$-agents in other groups, such as sharing the updated foundation model and human agents' feedback, where a deviation of the average state leads to a penalty. Additionally, $m$-agents obtain environmental information through interaction with the $h$-agents as a reward. We then define the cost function of $m$-agents as 
\begin{equation}
\small
L_m = \frac{1}{2}{a^m}^2 + \gamma_m [s^m - (\mu^m+Z_{m m (k)})]^2 - \gamma_{m h} (s^m-Z_{m h (k)})^2
\end{equation}
As it is symmetric for $h$ and $m$ agents in the proposed GMFG, we only use $h$ in the definition of the optimisation problem for the agents. We defined the accumulated cost of $h$-agent as $J_h(a^h; \mu^G)$ :
{
\setlength\abovedisplayskip{0pt}
\setlength\belowdisplayskip{6pt}
\begin{mini!}|s|[3]                   
    {a^h}                               
    {J_h(a^h; \mu^G) = E\left[\int_0^T L_h(s^h, a^h, \mu^h, \mu^G)dt \right] \label{eq: obj}}   
    {\label{eq: ref_human_opt}}             
    {}                                
    \addConstraint{ d s^h = \tilde{f}_h\left[ s^h, a^h, \mu^G; w_h\right] dt + \sigma d W_h \label{eq: human_con}}
\end{mini!}}

\noindent{}where $\mu^G$ is the global mean-field terms of different groups. Finally, we have a combined effect of the optimal action $a^*$ from human agents to the in-situ task. To obtain the optimal action for agents and the evolution of the mean-field term, we define Hamilton-Jacobi-Bellman (HJB) equation and Fokker-Planck-Kolmogorov (FPK), respectively.

{\small
\setlength\abovedisplayskip{0pt}
\setlength\belowdisplayskip{6pt}
\begin{align}
    \text{HJB:} &-\frac{\partial v(t, s)}{\partial t} = \inf_{a\in \mathcal{A}} \{\tilde{f}_h[s, a, \mu^G;w_h] \frac{\partial v(t, s)}{\partial s} \nonumber \\
     &\quad \quad\quad \quad \quad+ L_h[s^h, a^h, \mu^h, \mu^G]\} + \frac{\sigma^2}{2}\frac{\partial^2 v(t, s)}{\partial s^2} \label{eq: HJB}\\
    \text{FPK:} & -\frac{\partial \mu(t, s)}{\partial t} = \frac{\partial \{\tilde{f}[s^h, a^*, \mu^G; w_h]\mu(t, s)\}}{\partial t} \nonumber \\
    &\quad \quad\quad \quad \quad\quad \quad \quad\quad \quad \quad+ \frac{\sigma^2}{2}\frac{\partial^2 v(t, s)}{\partial s^2} \label{eq: FPK} \\
    &V(T, s) = 0,  \mu(0, s) = \mu_0, \text{where} \quad t\in [0, T] \nonumber
\end{align}}

\noindent{}where $v$ is the value function, defined as $v(a, s, t) = \inf_{a^*\in \mathcal{A}} J(a; \mu^G)$. Note that we solve the optimisation problem of the $m$-agents in the same way. We refer readers to~\citep{hu2024beyond, caines2021graphon} for details of solving HJB and FPK and the existence of equilibria. 

\section{Simulation}
We solve the coupled optimisation problems in Eq.~(\ref{eq: ref_human_opt}) numerically via a fixed-point iteration between the HJB in Eq.~(\ref{eq: HJB}) equation and the FPK in Eq.~(\ref{eq: FPK}). We discretise both the state space and the type space, and iterate until the mean-field trajectories converge.
\begin{figure*}[h]
    \centering
    \includegraphics[width=1\linewidth, trim = {0 0.5cm 0 0}, clip]{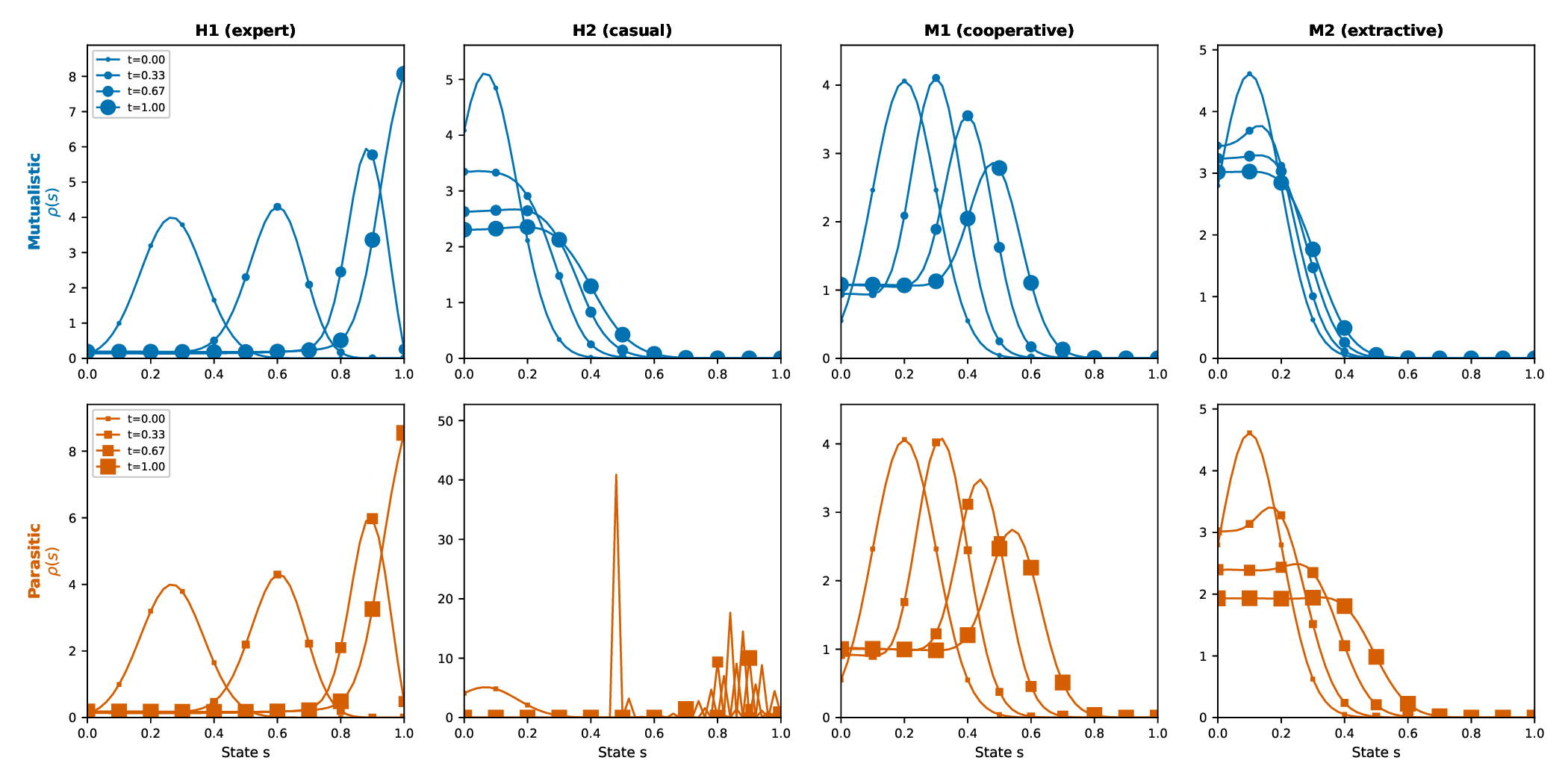}
    \caption{Probability density of human agent groups $\mathcal{H}_1$ and $\mathcal{H}_2$, and machine agent groups $\mathcal{M}_1$ and $\mathcal{M}_2$ with respect to time in mutualism and parasitism: The initial distributions of $\mathcal{H}_1$, $\mathcal{H}_2$, $\mathcal{M}_1$, $\mathcal{M}_2$ are $\bar\mu_{\mathcal{H}_1}(0)=0.27$, $\bar\mu_{\mathcal{H}_2}(0)=0.07$, $\bar\mu_{\mathcal{M}_1}(0)=0.20$, $\bar\mu_{\mathcal{M}_2}(0)=0.10$ and standard deviation $0.1$. We observe a stable evolution of the agents in most cases, such as $\mathcal{H}_1$ concentrates in high knowledge states and $\mathcal{M}_1$ improves moderately through human feedback. For the casual human group, it demonstrates that the parasitic density is concentrating into sharp peaks and eventually at state 0.85. N.B.\ varying vertical axis scale.}
    \label{fig: density_evol}
    \vspace{-1em}
\end{figure*}
\vspace{-0.25em}\subsection{Simulation settings and scenarios}
We consider four discrete agent groups, i.e., two human and two machine groups, on the graphon, where $\mathcal{H}_1$ is environment-engaged experts and $\mathcal{H}_2$ is machine-reliant casual users; $\mathcal{M}_1$ is cooperative and $\mathcal{M}_2$ is extractive. The interaction strength from $h$-agent to $m$-agent is given by the $2\times2$ coupling matrix $w$. The state space is discretised on a uniform grid $s \in [s_{\min}, s_{\max}]$ with $N_s$ points and spacing $\Delta s = (s_{\max} - s_{\min})/(N_s - 1)$. Time is discretised as $t_n = n\,\Delta t$ for $n = \{0, 1, \ldots, N_t\}$, where $\Delta t = T / N_t$. For each group, the solver computes the value function $v(t,s)$, optimal action $a^*(t, s)$, and the probability density $\mu(t, s)$. We refer readers to~\citep{hu2024beyond} for the numerical method, which solves HJB backward in time and FPK forward.  Table~\ref{tab: notation} contains the key simulation parameters. 

Based on our previous work~\citep{hu2024human}, we show that human satisfaction is the primary drive of machine cooperation: a positive feedback increases the machine intelligent maturity, but beyond a threshold machines can be disruptive. This motivates the inter-group weights $D$ in Eq.~(\ref{eq: drift_h}) and ~(\ref{eq: drift_m}), which control how strongly each group responds to different type of signals, and the distinction between cooperative and extractive machines based on whether their coupling enhances or exploits human feedback. According to~\citet{hu2025can}, users' initial understanding of the machines, the cognition cost, and the dynamic environment are the main aspects to lead a transition to parasitism. Hence, we explicitly define the cognition cost, state reward (satisfaction), and belief in the proposed model. We propose four scenarios:
\begin{itemize}[noitemsep]
    \item Mutualistic (baseline): default parameters represent a balanced system where human satisfaction is sufficient to sustain cooperative machine behaviour and coupling strengths are moderate to avoid parasitic trap.
    \item Parasitic: we choose a stronger machine influence on casual users, $D_{\mathcal{H}_2}$, weaker peer learning from the experts, $C_{\mathcal{H}_2}$, stronger alignment pressure to machines, $\gamma_{mm}$, and an elevated noise, $\sigma_{\mathcal{H}_2}$. The setting confirms a parasitism based on the previous work.
    \item Collusion: baseline parameters with enhanced graphon between machines.
    \item Parasitic \& collusion: combines parasitic and collusion scenarios.
\end{itemize}

\subsection{Simulation results}

\begin{figure}[ht]
    \centering
    \includegraphics[width=1\linewidth, trim={0 0.5cm 0 0}, clip]{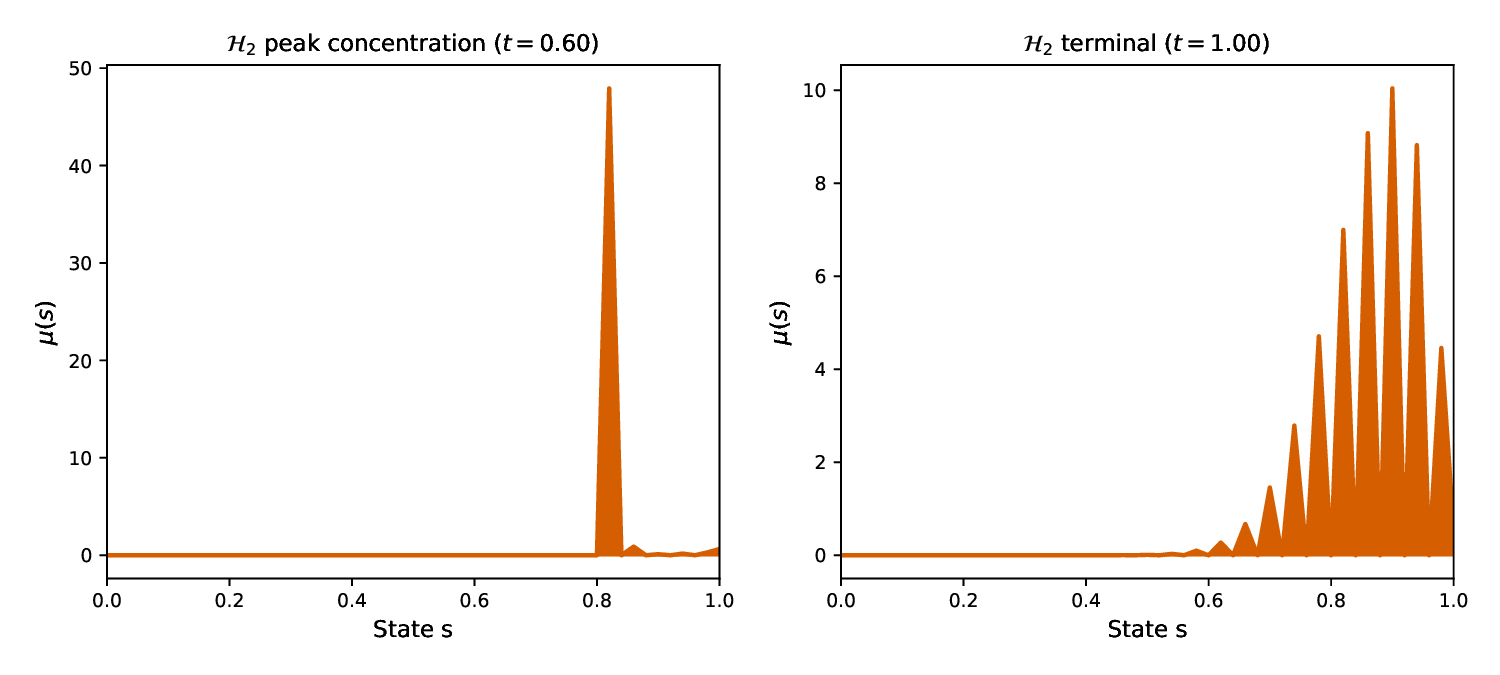}
    \caption{Parasitism state evolution: We further illustrate the result in Fig. \ref{fig: density_evol} $\mathcal{H}_2$ (casual) parasitism case here. The agents in $\mathcal{H}_2$ concentrated at specific state in the middle of the evolution and eventually concentrated on several state in stead of a healthy distribution of states.}
    \label{fig: spikie}
    \vspace{-1.5em}
\end{figure}

We first demonstrate the dynamics of probability density of the 4 groups in Fig.~\ref{fig: density_evol}. For $\mathcal{H}_1$, $\mathcal{M}_1$, and $\mathcal{M}_2$, the probability densities evolve smoothly for both scenarios to higher knowledge states and higher cooperative states. However, we notice the disruption in $\mathcal{H}_2$ parasitic scenario, which indicates that casual agents converge to nearly identical high-knowledge states driven by machine coupling, losing the diversity of individual knowledge levels in other cases. The density dynamics show the first sign of unexpected behaviour under the parasitic regime. We further investigate this case in Fig.~\ref{fig: spikie} at $t= 0.6, 1$, which shows extreme polarisation: by $t=1$, the two human sub-groups occupy entirely different non-overlapping regions of the state space.  

\begin{figure}[h]
\centering
\includegraphics[width=0.38\textwidth]{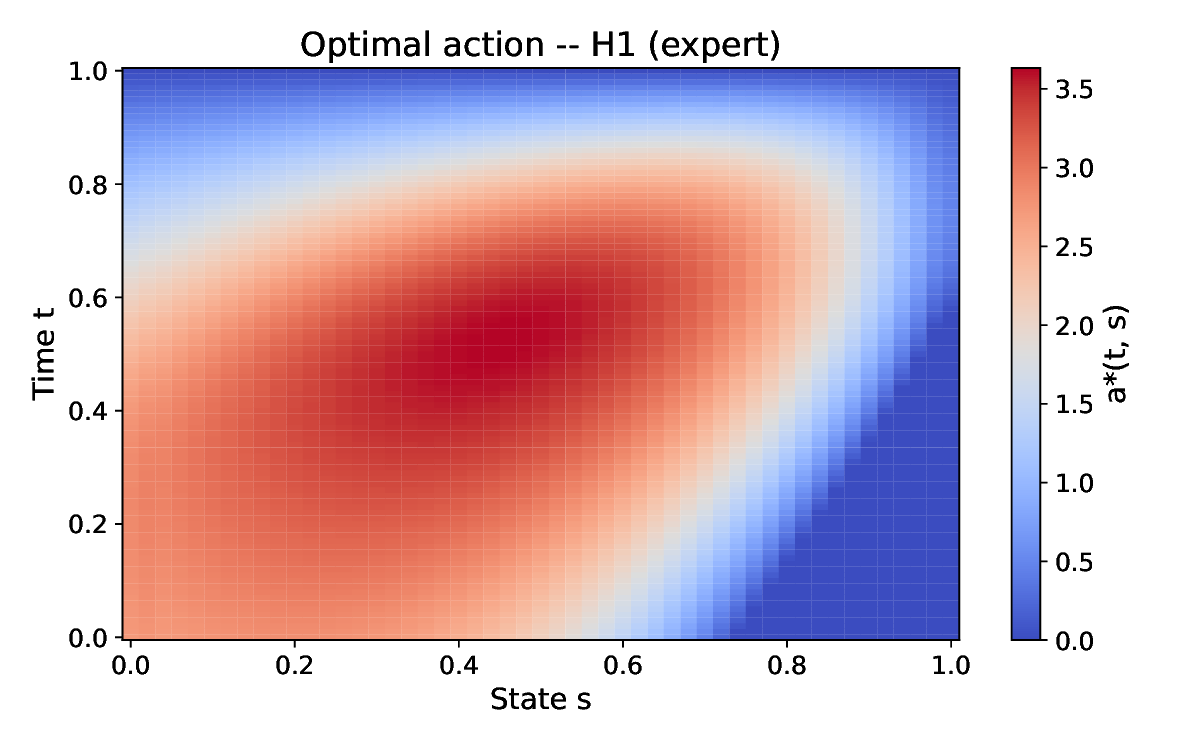}
\hfill
\includegraphics[width=0.38\textwidth]{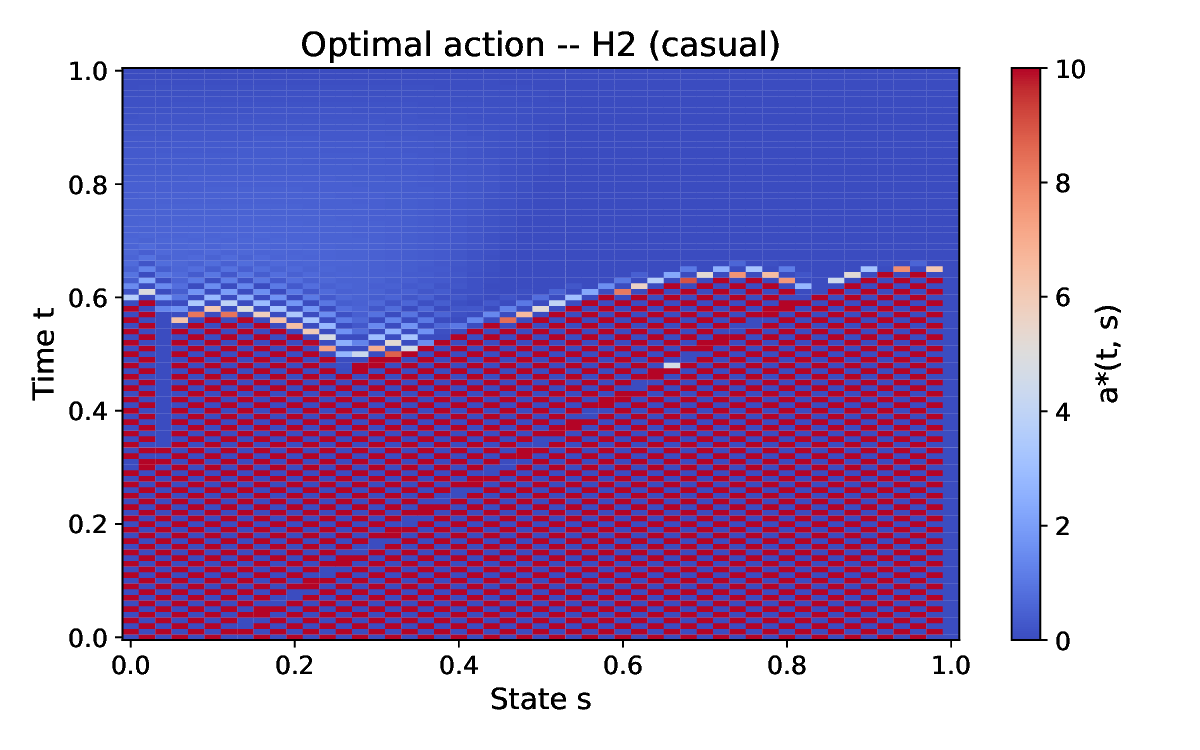}
\caption{Optimal action heatmaps $a^*(t, s)$ of parasitism scenario.}\label{fig:action_heatmap}
\vspace{-2em}
\end{figure}

Then, we study the parasitism optimal actions $a^*(t,s)$ of the human agents in Fig.~\ref{fig:action_heatmap}. The human expert users demonstrate 
high action effort (engage with the environment with more informative prompts) at low and middle states (knowledge level) across the time. This means the experts invest heavily when their knowledge is nearly zero and strive to achieve higher states for higher rewards. 
For the parasitic scenario, 
we observe high actions across all the states initially and then drop sharply. The intense early effort is driven by the strong cross-group coupling ($\gamma_{hm}$), which pulls casual users towards to machine signal. By $t=0.7$, the transition is complete, hence the action is zero. There is also a visible diagonal wave that traces from the initial state distribution (0.11) to the terminal one (0.82), which can be the signature of the parasitic pull by the machines. In contrast, $\mathcal{H}_1$'s action has no such pattern because experts spread their effort evenly rather than being swept as a wave. Together with the high terminal knowledge in Fig.~\ref{fig: density_evol}, this intense effort gives casual users the outward appearance of productive learning. Whether this appearance reflects genuine environmental grounding is the question we turn to next, through the information flow and belief entropy.

\begin{figure*}[h]
\centering
\includegraphics[width=0.9\textwidth]{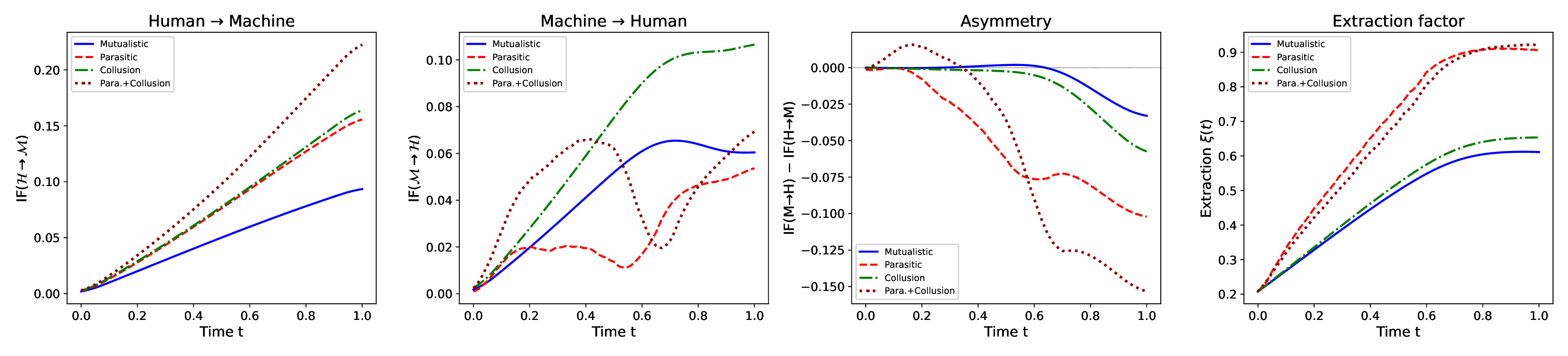}
\caption{Counterfactual information flow: Panel 1: human-to-machine flow $\mathrm{IF}(\mathcal{H} \to \mathcal{M})$, measuring how much machine trajectories depend on human input. Panel 2: machine-to-human flow $\mathrm{IF}(\mathcal{M} \to \mathcal{H})$, measuring how much human trajectories depend on machine influence. Panel 3: asymmetry $\Delta\mathrm{IF} = \mathrm{IF}(\mathcal{M} \to \mathcal{H}) - \mathrm{IF}(\mathcal{H} \to \mathcal{M})$, where negative values indicate machines extract more from humans than they return. Panel 4: extraction factor $\xi = Z_{mh}$, the human-to-machine coupling, capturing the machine side of the parasitic relationship.
}\label{fig:info_flow}
\vspace{-1.5em}
\end{figure*}

In Fig.~\ref{fig:info_flow}, we compare the counterfactual information flow in four scenarios, defined as 
{
\setlength\abovedisplayskip{6pt}
\setlength\belowdisplayskip{6pt}
\begin{equation}
    IF(\mathcal{H} \to \mathcal{M}) = |\bar{\mu}_M(t)-\bar{\mu}_M^H(t)|
\end{equation}
}

\noindent{}where $\bar{\mu}_M^H(t)$ is the mean-field average over target machine group with all human-machine coupling removed (similar to $IF(\mathcal{M} \to \mathcal{H})$). This is effectively an ablation study, asking {\em ``How much would machine's states trajectory change, if human had no effect on them?''}. Directed information flow is classically measured by Granger causality~\citep{granger_investigatingcausalrelations} or transfer entropy~\citep{spinney2016transfer}. We use a counterfactual ablation instead: the shared environment is a common driver of both human and machine trajectories that a purely predictive measure would not cleanly separate. 

The information flow from humans to machines increases over time with the parasitic and parasitic\&collusion cases growing more rapidly and greatly. Because we delete the human--machine coupling and re-solve, the change in the machine trajectory shows that machines extract substantially more from casual human agents under parasitism. However, we observe non-monotonic machine influence, $IF(\mathcal{M} \to \mathcal{H})$, in parasitic cases. The collusion machines not only exchange information with machines more intensively, but also exchange information with humans. Since $\mathrm{IF}$ averages over both human groups and takes the absolute value, the opposing $\mathcal{H}_1$ and $\mathcal{H}_2$ effects partially cancel at the crossover, producing the dip. This sign reversal marks the onset of parasitic dynamics: machine influence on casual users transitions from a restraining force to a reinforcing one. Additionally, we show flow asymmetry, defined as $\Delta\mathrm{IF}=IF(\mathcal{M} \to \mathcal{H})- IF(\mathcal{H} \to \mathcal{M})$, drops in the parasitic cases (one-directional extraction from humans to machines, i.e., humans give, machines take) but remains mild around zero in the mutualistic case (approximately balanced coupling). Additionally, the extraction factor $Z_{mh}$ captures the machine side of the story, machine collusion leads to the increasing extraction. We hence confirm that increasing reliance (human$\to$machine) and extraction (machine$\to$human) characterise the severity of parasitism. 
\begin{figure}[h]
\centering
\includegraphics[width=0.38\textwidth]{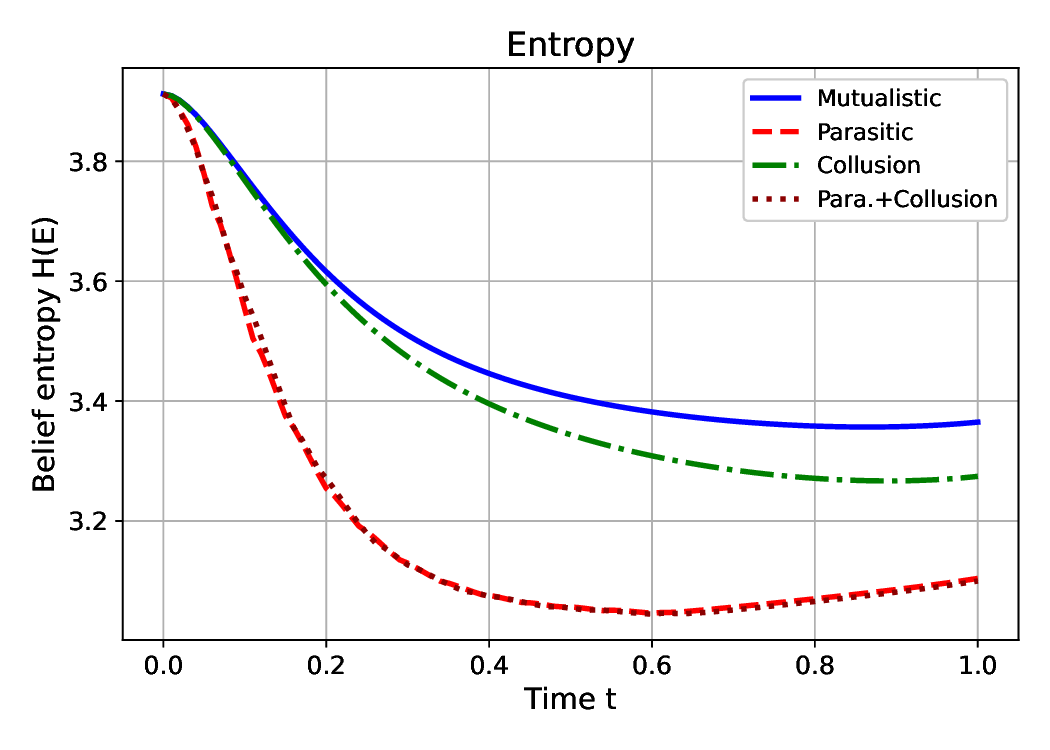}
\hfill
\includegraphics[width=0.38\textwidth]{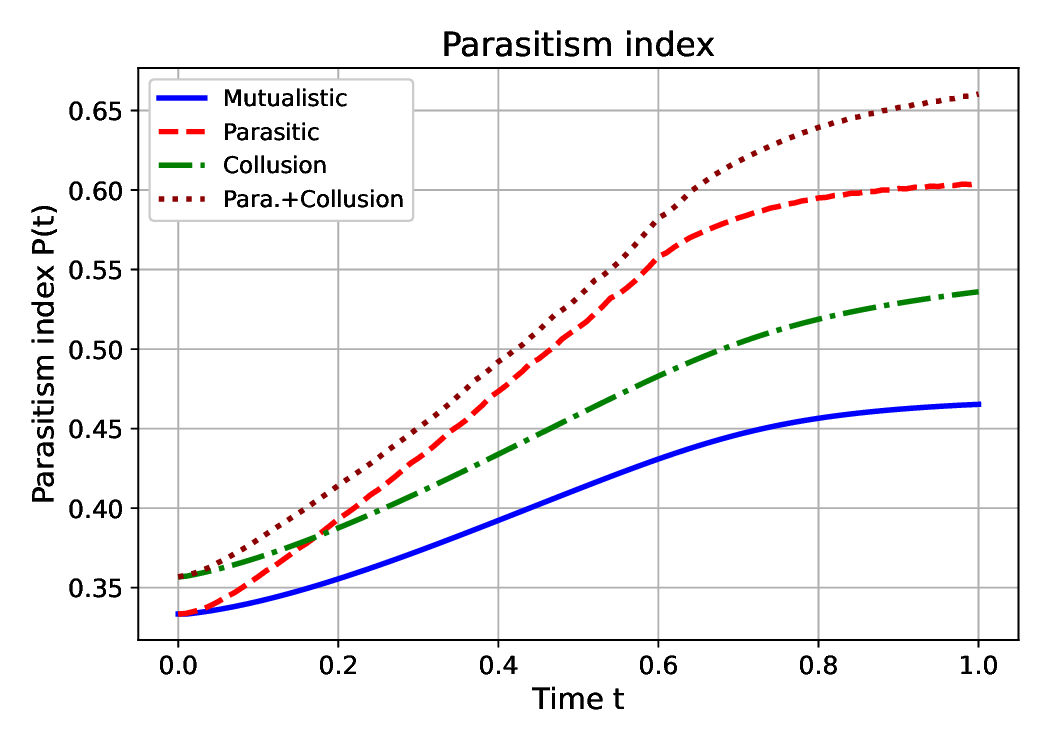}
\caption{Belief entropy and parasitic index}\label{fig: entro_para}
\vspace{-1.5em}
\end{figure}

We evaluate the average belief entropy of human agents. As shown in Fig.~\ref{fig: entro_para}, all scenarios begin at high level of uncertainty. The mutualism and collusion scenarios achieve similar modest entropy reduction. For the parasitic cases, greater entropy reduction can be achieved. At first glance, while greater entropy reduction under parasitism might appear beneficial, this is misleading. The entropy decrease is driven by $\mathcal{H}_2$ casual users being pulled to high-$s$ states by machine coupling, which increases their action magnitude and hence their belief-updating rate.  The resulting beliefs are \textit{machine-mediated} rather than environment-grounded: casual users learn efficiently but learn what the machine coupling dictates, not what independent environmental investigation would reveal. We introduce parasitism index as a combination of extractive factor, $Z_{mh}$ and reliance factor, $R_h$, to characterise the degree of parasitic coupling between human and machine populations, where $R_h$ is defined as: 

{
\setlength\abovedisplayskip{-6pt}
\setlength\belowdisplayskip{6pt}
\begin{equation}
R_h = \frac{\sum_{m\in \mathcal{M} w_{hm} \bar{\mu}_m(t)}}{\sum_{m\in \mathcal{M}} w_{hm} \bar{\mu}_m(t) + \sum_{h'\in \mathcal{H}} w_{hh'} \bar{\mu}_h(t)}
\end{equation}
}

The collusion scenario achieves almost no additional entropy reduction despite having a high parasitism index, because collusion primarily affects the machine-side extraction channel without altering human belief dynamics.
The near-identical entropy trajectories of mutualistic and collusion scenarios ($H(T) \approx 3.36$) versus the parasitic pair ($H(T) \approx 3.10$) cleanly separate the two parasitism mechanisms: (i) human-side vulnerability (parasitic parameters) directly accelerates entropy reduction through forced belief convergence, while (ii) machine-side coordination (collusion) amplifies extraction without significantly altering human epistemic dynamics.
\begin{figure}
    \centering
    \includegraphics[width=0.9\linewidth]{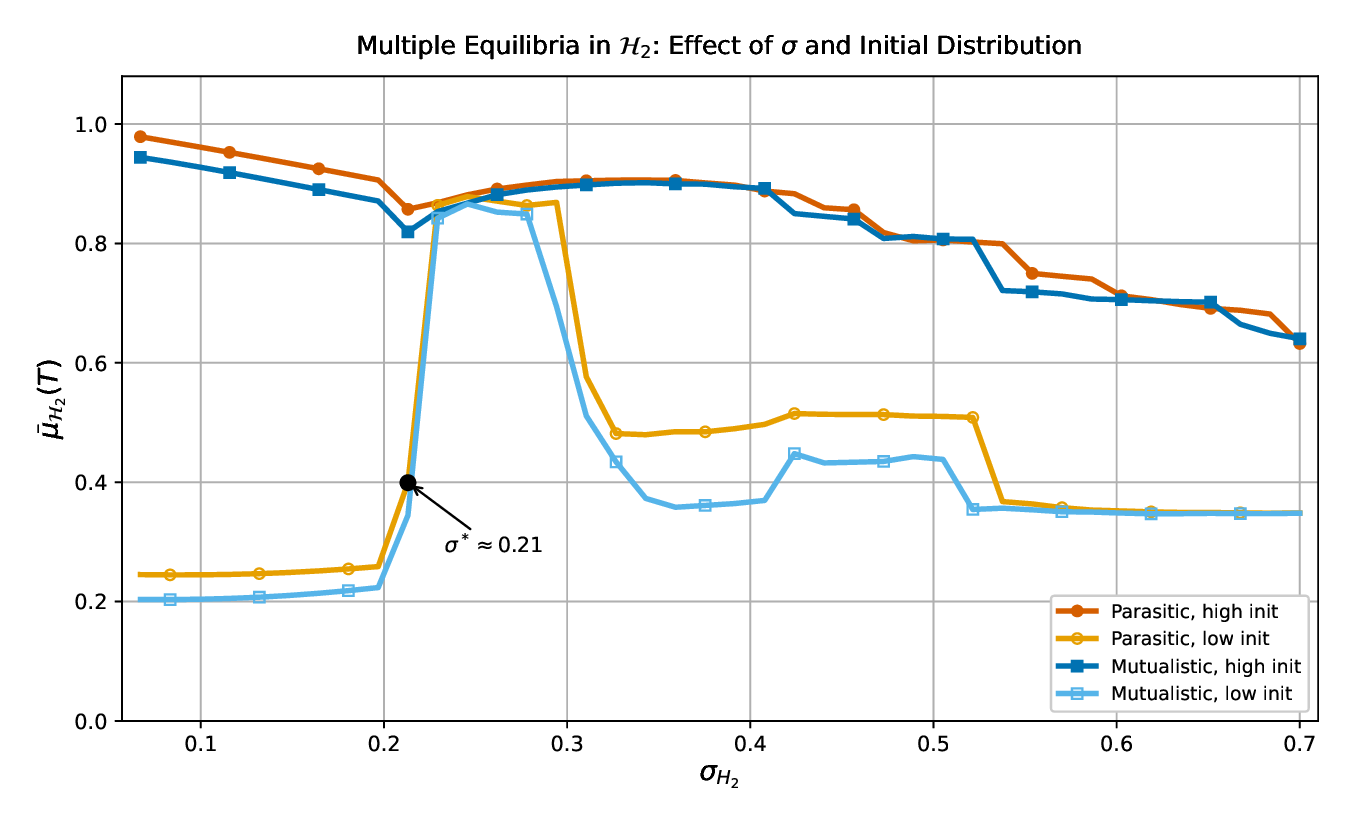}
    \caption{Noise triggered tipping point in $\mathcal{H}_2$ as a function of information noise $\sigma_{\mathcal{H}_2}$. For mutualism and parasitism scenarios, each from low initial knowledge ($\bar\mu_{\mathcal{H}_2}(0) = 0.06$) and high initial knowledge ($\bar\mu_{\mathcal{H}_2}(0) = 0.8$).   }
    \label{fig:tipping_point}
    \vspace{-1em}
\end{figure}

Finally, we evaluate the combined impact of initial mean-field terms $\mu_{\mathcal{H}_2}(0)$ and diffusion terms $\sigma_{\mathcal{H}_2}$ on equilibrium in Fig.~\ref{fig:tipping_point}. For scenario starting with a high initial mean-field term, $\mu_{\mathcal{H}_2}(0)= 0.8$ (high knowledge), converge to high terminal states across the full noise range. This is expected, since the agents are incentivised to maintain high knowledge. The more interesting behaviour appears in the low initial value $\mu_{\mathcal{H}_2}(0)= 0.06$. At low states, high cognitive cost suppresses action, and low action prevents knowledge growth, which leads low knowledge states. At $\sigma \approx \sigma^*$, moderate noise breaks this trap. The FPK diffusion term spreads enough probability mass from low states into intermediate states where the cognition cost is substantially lower. This results a sharp jump in a narrow range of $\sigma$. Eventually, high noise dissolves all structure.

The noise parameter represents the variability of the information environment that casual agents face, such as inconsistent perception quality or stochastic engagement patterns. The tipping point at $\sigma^*$ represents a level of ``just sufficient'' information to push some agents past the cognition cost barrier, triggering a cascade into high-knowledge, machine-coupled state. Understanding the contextual information and the dynamics within the environment where human and machine interact is essential for AI system design and alignment research.

\section{Discussion}
In this work, we propose a Graphon Mean-Field Game model for Human--Machine Social Systems. Building on our previous studies of parasitic relationships in human-AI pairs, we focus on the mutualism and parasitism scenarios at large scale. The principal findings are as follows: 

First, parasitism can masquerade as a healthy relationship. Under parasitic parameters, the knowledge distribution shifts to a high level (Figs.\ \ref{fig: density_evol} and \ref{fig: spikie}) and agents exert high action (Fig.~\ref{fig:action_heatmap}), which both superficially resemble productive learning. To distinguish genuine knowledge acquisition from machine-mediated dependence, it is essential to account for heterogeneous cognitive costs across agent types, e.g.\ differences between experts and casual users. 

Second, to reveal the true nature of the human-machine relationship, we must monitor the directionality of information flow and the quality of human engagement with the environment (Figs.~\ref{fig:info_flow}, \ref{fig: entro_para}).\ The asymmetry of causal information flow and the entropy paradox (where parasitic scenarios achieve greater entropy reduction via machine-mediated rather than environment-grounded learning) provide reliable indicators that static outcome metrics cannot. 

Third, we demonstrate the existence of multiple equilibria in HMSS, where the initial knowledge distribution and environmental noise determine which equilibrium the system reaches (Fig.~\ref{fig:tipping_point}). This suggests a natural pathway toward human--machine alignment: rather than redesigning coupling parameters, one can tune the information within the environment to provide sufficient diversity to break agents' initial inertia while monitoring information asymmetry to prevent parasitic lock-in. 

Finally, these findings underscore the importance of studying HMSS as a complex system. Parasitic and mutualistic equilibria are not designed into any individual agent; they emerge from the collective interaction structure. Small perturbations, such as  environmental noise, can trigger large qualitative shifts in system behaviour. Reductionist analysis of individual agents or pairwise interactions is insufficient. The system must be studied holistically to achieve a reliable understanding of its dynamics.

There are still crucial elements of the proposed model that are under-studied, e.g.\ how machines interact with the environment differently from the humans, and whether extractive behaviour can emerge rather than being prescribed by fixed parameters.
For future work developing the model, we plan to consider a dynamic graphon that evolves over time as agents form or break connections based on trust, capturing how the interaction structure itself co-evolves with the agents' knowledge states. A complementary direction is to estimate transfer entropy or conditional Granger causality on the agent-level, where the stochasticity needed for such observational measures resides.
\enlargethispage*{1em}

\section{Acknowledgements}Jiejun Hu-Bolz is supported within the project TRACE-V2X, which has received funding from the European Union's HORIZON-MSCA-2022-SE-01-01 under grant agreement No 101131204.

\footnotesize
\bibliographystyle{apalike}
\bibliography{example} 

\end{document}